\RequirePackage{fix-cm}
\documentclass[smallextended]{svjour3}       % onecolumn (second format)
\smartqed  % flush right qed marks, e.g. at end of proof
\usepackage{graphicx}
\usepackage{amssymb,amsmath}
\begin{document}

\title{Adjoint-based Data
  Assimilation of an Epidemiology Model for the
  Covid-19 Pandemic in 2020}
%\subtitle{doi://10.5281/zenodo.3732292}

\titlerunning{Covid-19 Data Assimilation}        % if too long for running head

\author{J\"orn Lothar Sesterhenn}


\institute{J.L Sesterhenn \at
              Universität Bayreuth \\
              Tel.: +49 (0) 921/55-7261\\
              \email{joern.sesterhenn@uni-bayreuth.de}\\
              ORCID ID: 0000-0003-1309-2306
}

\date{Received: date / Accepted: date / \\{doi://10.5281/zenodo.3732292}}

\maketitle

\begin{abstract}
  Data assimilation is used to optimally fit a classical epidemiology
  model to the Johns Hopkins data of the Covid-19 pandemic. The
  optimisation is based on the confirmed cases and confirmed
  deaths. This is the only data available with reasonable
  accuracy. Infection and recovery rates can be infered from the model
  as well as the model parameters.  The parameters can be linked with
  government actions or events like the end of the holiday
  season. Based on this numbers predictions for the future can be made
  and control targets specified.

  With other words:

  \textbf{We look for a solution to a given model which fits the given
    data in an optimal sense. Having that  solution, we have all
    parameters.}
\end{abstract}


\section{Introduction}
\label{intro}
The Covid-19 outbreak is threatening our health and lives. Without
intervention, the numbers of infected people worldwide will rise to
millions and leave hundreds of thousands of casualties behind, prior
to settling down because the spread is stopped due to the fact that
the virus sees no susceptible victims in a large enough number to
further spread. This moment is reached, when the spreading rate
\(\beta\) and recovery rate \(\gamma\) have a specific ratio or
the number of susceptible people has become low enough. Waiting for
the latter meants a long time and a high death toll. There is no way
of controling \(\gamma\), only \(\beta\) depends on the behaviour of
the people and is thus amendable to our intervention. In lack of
either a vaccine or a cure this is the only amendable
parameter. \cite{FergusonEtal2020} have studied how this can be
done. This currently goes by the term \emph{flattening the curve}. In
order to make a valid assessment of actions taken in the past or to
make reliable predictions for the future, reliable numbers are needed,
but hard to get. Even if specific numbers were available, the
presently infected or recovered people can only be infered by means of
a model.

A good data source for global data is the github repository of the Johns
Hopkins project Systems Science and Engineering \cite{CSSE2020}. Initially they
distributed confirmed cases (C), recovered cases (R) and deaths (D). Due
to the notorious unreliability of (R) this datum was discontinued. The
task now is to get from this data good estimates of the real numbers as
well as the hidden states. Data assimilation is capable of recovering
the full state of a model from partial information.

In the sequel data assimilation will be applied to the Covid-19 data.
The method can be summarised as a mathematical method to get out of a set
of empirical data \(C_e,D_e\) plus a model \({\cal M}\) the full state
\(S,I,R,D\) and model parameters \(\beta, \gamma, \delta\) such that the
data is optimally described:
\[
  C_e,D_e,{\cal M} \longrightarrow S,I,R,D,\beta, \gamma, \delta.\] In
addition to the variables introduced above, \(S\) is the number of
susceptible individuals in the population. It is effectively this
number of not immune people, which stops the outbreak when it falls
low enough in the natural case.

The result will obviously not be better that the model, but even with
a classical epidemiology model they are much better than using
statistical means of getting the information. At this point, there is
ample room for improvement and the author, being a physisist,
solicitates colleagues with better knowledge about epidemology models
to contact him.
\section{Method}
\label{sec:1}
\subsection{Adjoint Method}
Lemke et al. \cite{LemkeCaiReissPitschSesterhenn2018} presented a method for a
sensitivity analysis of complex reaction mechanisms based on an
adjoint approach. The basic structure of the equations is similar and
the method is applicable to epidemics.

Here a very brief sketch is given to ease understanding of the procedure
without all the details from the previous publication. The governing
equations have the form \[{\cal M}=\partial_t q -
f(q,\alpha)=0.\] While in the cited paper different chemical species are
considered, we use the identical governing form for a model adopted from
the classical SIR model \cite{Hethcote2000} modified for a mortal class.
Imagine the number of susceptible, infected, recovered and dead people
\(q=(S,I,R,D)^T\) as being the chemical species in a reaction, \(\alpha\)
are the parameters of the model and in our case they are taken to be
\(\alpha=(\beta, \gamma, \delta)^T\) for the contact, recovery and
mortality rate of the epidemic. The model was augmented by the number of
confirmed deaths \(D\), because that is known from the data, whereas the
recovered class \(R\) is uncertain an no longer provided by the source
since 2020-03-23.

\subsection{Epidemology Model}
\label{epidemology-model}

The model then reads \begin{eqnarray}
                       \partial_t S &=& \beta IS/N\\
                       \partial_t I &=& -\beta IS/N -(\gamma+\delta) I \\
                       \partial_t R &=& \gamma I\\
                       \partial_t D &=& \delta I
                     \end{eqnarray}
with \(N=S+I+R+D\)
being the total population including the dead. The model can easily be
augmented to accommodate several other classes like age classes or local
distributions representing several countries. This set of equations has to be
solved with an initial state \(q_0=(S_0,I_0,R_0,D_0)^T\) subject
minimising a constraint
\begin{equation}
  {\cal J}=\int (C^Tq-(C^Tq)_e)^2 dt.
\end{equation}
\((C^Tq)_e\) is the empirical data towards which the assimilation is
performed. It is known only as a a compound and can not be disentangled
into its components. This is indicated by the brackets
\(\left(\cdot\right)_e\). \(C\) describes that compound in form of an
observation matrix.

With other words:

\textbf{we look for a solution to
a given model which fits the given data in an optimal sense. Having that
solution, we have all parameters.}

The data was taken from the Johns Hopkins github repository Systems
Science and Engineering (2020). We observe the total confirmed cases
\(I+S+D\) and the confirmed deaths \(D\). Thus we have
\[
  C=
  \begin{pmatrix}
    0&1&1&1\\
    0&0&0&1
  \end{pmatrix}.\]

Note, that it is
possible to weight each row differently, in case we rely better on one
source or the other. This is not done in this article.

The augmented Lagrangian to be minimised reads \[{\cal L} = {\cal
J} + q^* {\cal M}.\] \(q^*\) is a lagrangian multiplier which will be
the adjoint state vector. The trick is, that we added Zero for any value
of \(q^*\) since \({\cal M}=0\).

Linearisation of \(\cal M\) and \(\cal J\) yields after some steps,
(\cite{LemkeCaiReissPitschSesterhenn2018}), an adjoint equation to be solved:
\[\partial_t q^* = - \underbrace{\left(\frac{\partial f}{\partial
q}\right)^T}_{A^T} q^* - \underbrace{((C^Tq - (C^Tq)_e)^T C^T)^T}_g\] or
in short \begin{equation} \partial_t q^* = - A^T q^* - g.
\end{equation}

This is an equation for a particular \(q^*\). Rearranging the terms,
it turns out that our choice \(q^*\) is the key to assess changes in
model parameters:

\[ \int q^{*T}\frac{\partial f}{\partial \alpha} dt
\approx\frac{\partial {\cal J}}{\partial \alpha}.\]

This expression is called the \textbf{sensitivity} with respect to
changes in the control parameters. This is, what tells us which change
to the governing parameters are important and which not. They can be
interpreted as a the direction in which the cost functional gets worse
if the governing parameters \(\alpha\) are changed. In order to get
closer to the data, we need to follow it's negative direction. In
figurative terms, it directs us to where it goes up or down on a
mountain, and we need to follow the route down into the valley in our
case.

\section{Results and Discussion}\label{results-and-discussion}

In the sequel several data-sets are studied individually. China was the
origin of the pandemic and has been able to bring the outbreak under
control. We first study shortly the convergence towards the assimilated
state and then discuss the full states and the underlying epidemic
parameters.

\subsection{Convergence}\label{convergence}

\begin{figure}
\centering
\includegraphics[width=0.7\textwidth]{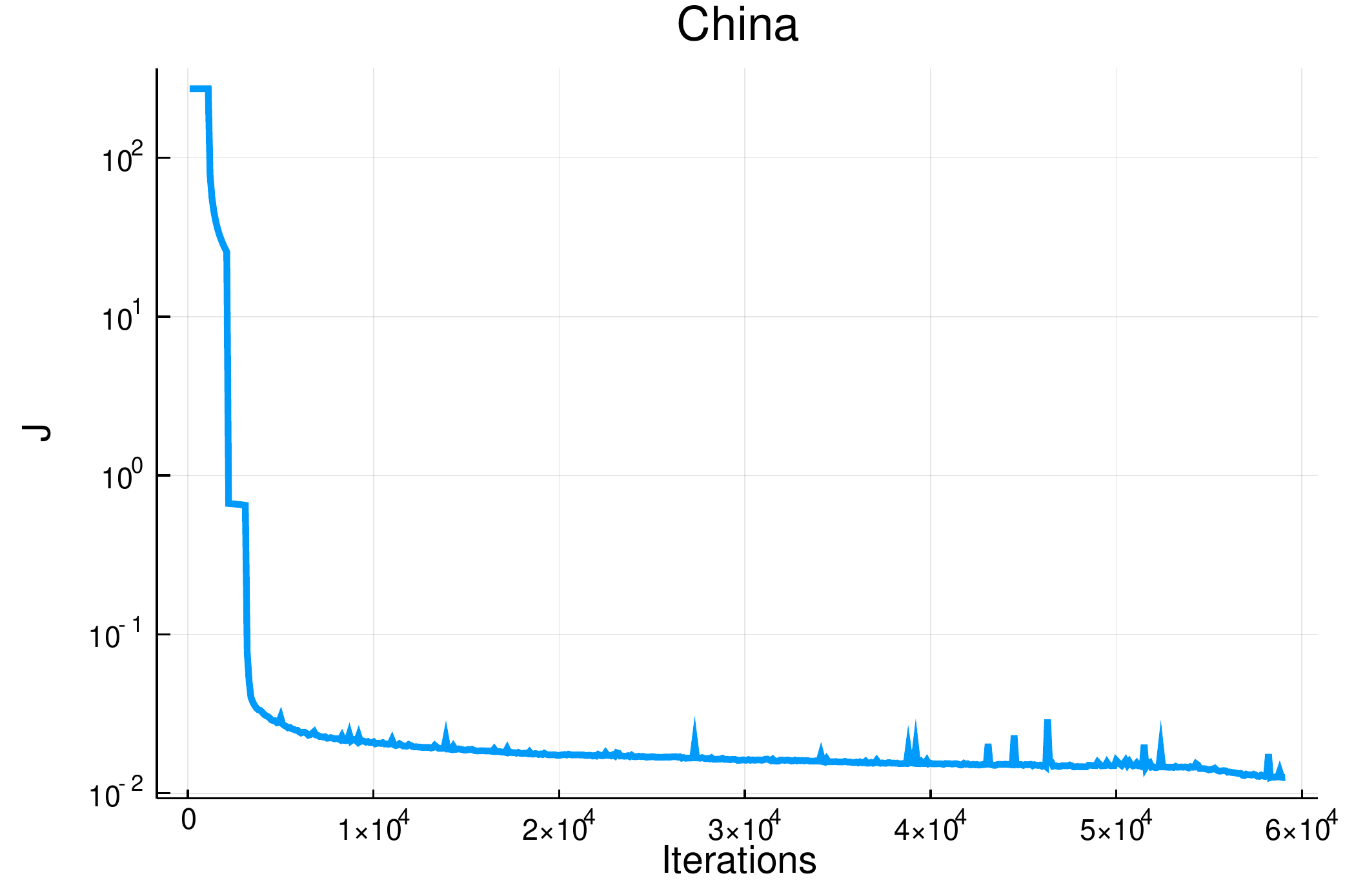}
\caption{Exemplary Convergence history for the case of China.}
\end{figure}

Convergence is measured as the norm of the discrepancy between data
and simulation, normalised by the norm of the data:
\[\frac{\| C^Tq-(C^Tq)_e\|_2}{\|(C^Tq)_e\|_2}.\] The calculation was
started with an arbitrary value of
\((\beta, \gamma,\delta)=(1/4,1/7,0)\). This leads to an exorbitant
growth in the shown case of China and correspondingly the norm is
initially very high as compared to the provided data. Within a few
iterations it comes down to approximately 7\% and then slowly
converges until the computation is stopped at an agreement of 2\%.

This value is sufficient to match the development of the pandemic
well. The start, when numbers are exceedingly small with One or Two
infected is not covered.

The present speed of convergence is not satisfactory when considering
large scale fluid-dynamical problems, which is, what the author
usually deals with, but since the present calculations run within less
an hour on a standard laptop, no attempt was made to improve
convergence in addition to what a brute force linesearch provides.

This needs to be improved upon, especially, when the onset of the
outbreak is to be studied.

\subsection{Case Studies}\label{convergence}

Several case studies will be presented: China, Italy, Germany,
Bavaria, the United States and the United Kingdom.  Cina has been the
first Country, where we can observe the full outbreak and decay. Italy
was the first European country affected an the first to take action.
Germany including Bavaria reacted next, and US and UK are very early
in the development.

\subsubsection{China}\label{china}

\begin{figure}
\centering
\includegraphics[width=0.75\textwidth]{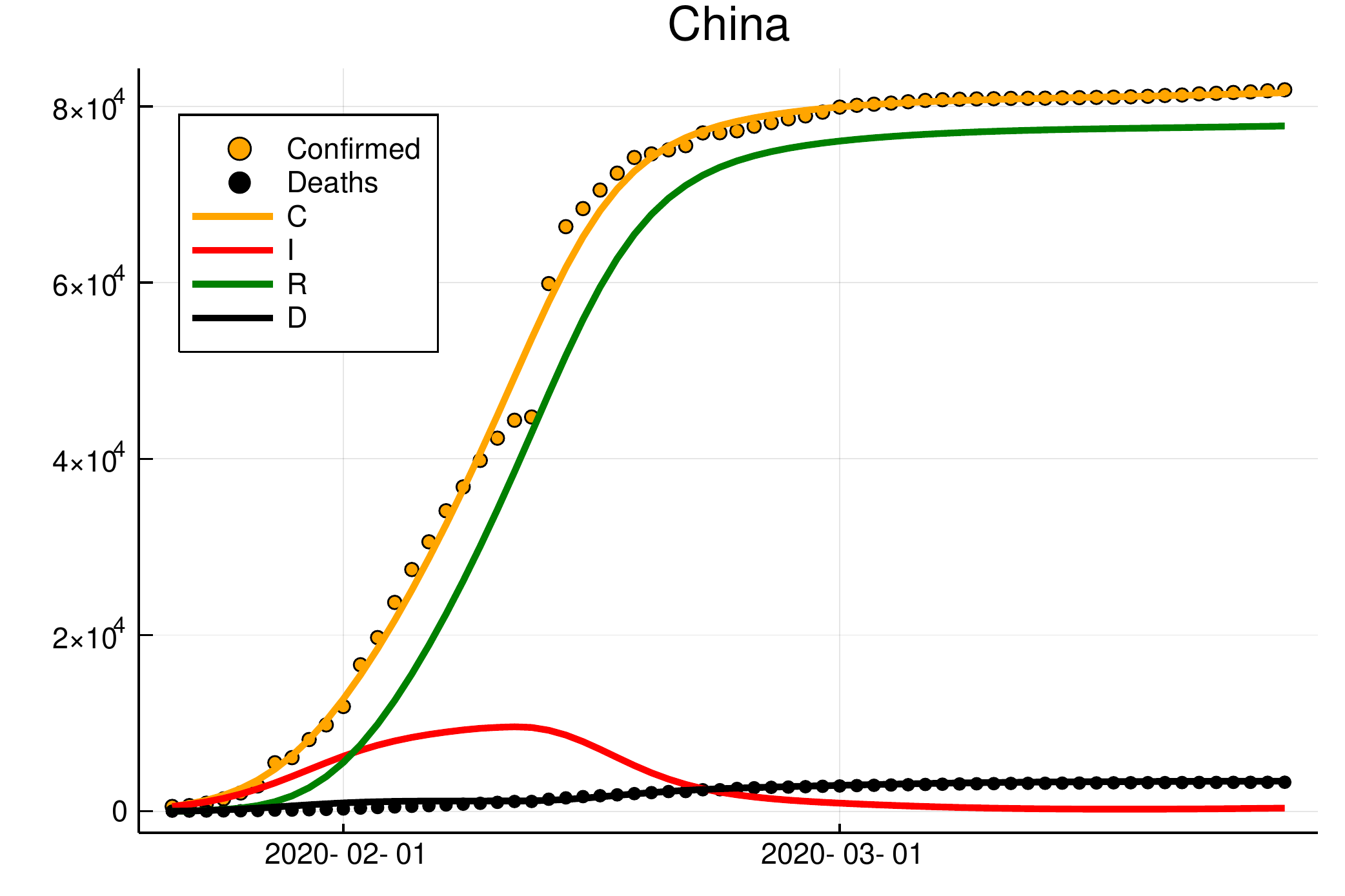}
\caption{Data from China (dots) and assimilation (lines)}
\end{figure}

The data for China in the Johns Hopkins data starts on January 22 and is
assimilated until present. Reported are the confirmed cases, which
represent the sum of infected, recovered and dead people, \(C=I+R+D\)
and deaths \(D\). After the first three weeks, there is an enormous jump
in the data which is likely due to a change in information policy of the
Chinese government or a change in how cases are reported. This
corresponds in a rise in presumably infected people. Unfortunately this
happened right before goverment action on February 26th.

\begin{figure}
\centering
\includegraphics[width=0.75\textwidth]{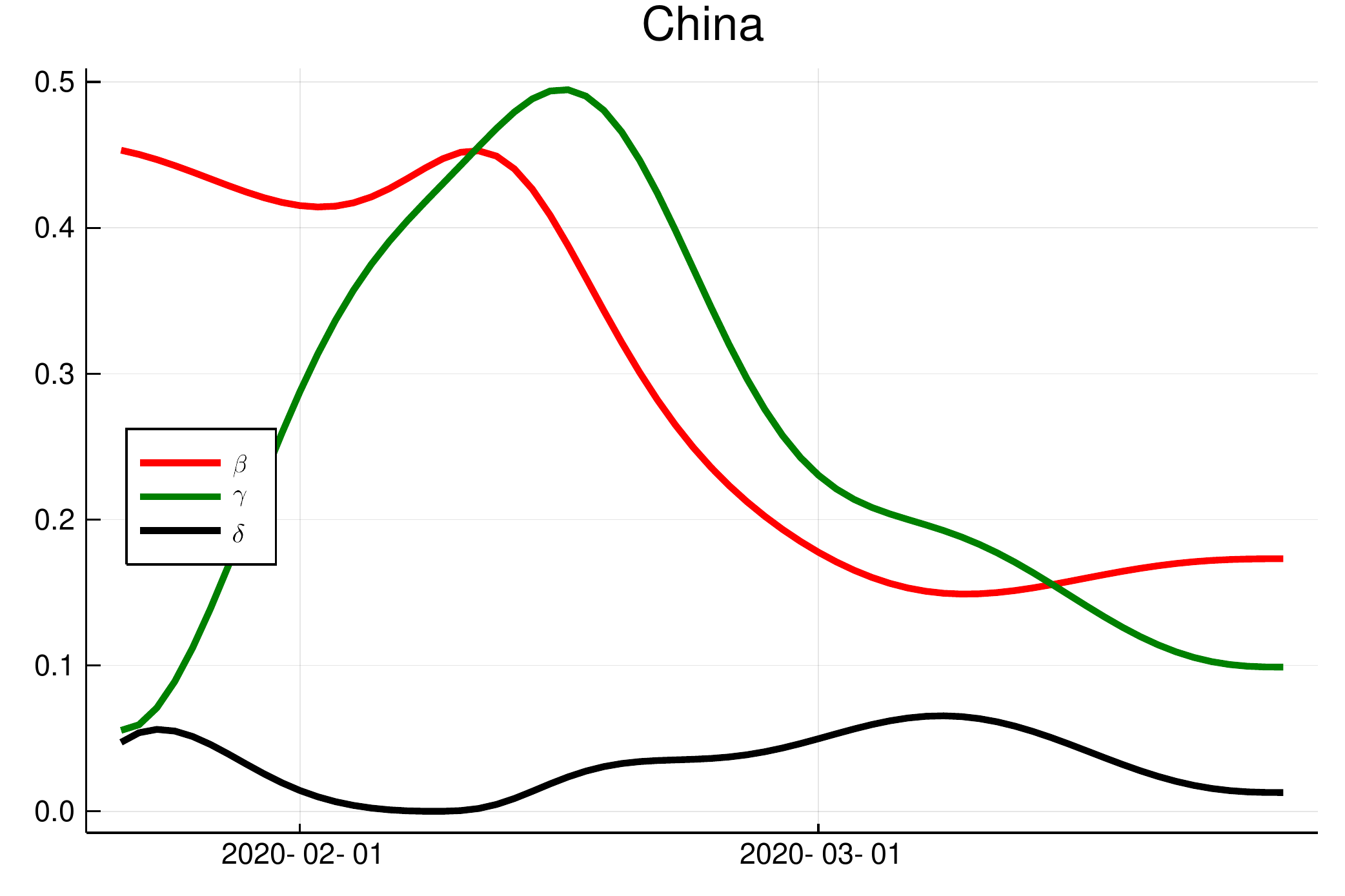}
\caption{Model parameters \(\beta, \gamma, \delta\) for China}
\end{figure}

Better than the full recovered state, the parameters
\(\beta(t), \gamma(t),\delta(t)\) show the development of the crisis,
and how effective were government activities. The dependence on time was
explicitly noted in the expression, to indicate that these parameters
are not constants, but are fitted to match the data.

On January 26, China closed Universities, prescribed remote working
measures and installed several other measures to contain the spreading
of the virus. These measures seem not to be effective until a week later
or obscured by the jump in reported confirmed numbers right at the same
time.

The transmission rate \(\beta\) then shows a drop from \(\beta=0.45\) to
\(\beta=0.16\) only from February 10 until March 1. This stopped the
outbreak. Also seen in the figure is the spurious trough in death rate
\(\delta\) at the end of January. This is
likey due to the spurious large number of new infects the algorithm
tries to compensate.

It is also visible that the infection rate \(\beta\) rises slightly
again since the second week of March.

\subsubsection{Italy}\label{italy}

\begin{figure}
\centering
\includegraphics[width=0.75\textwidth]{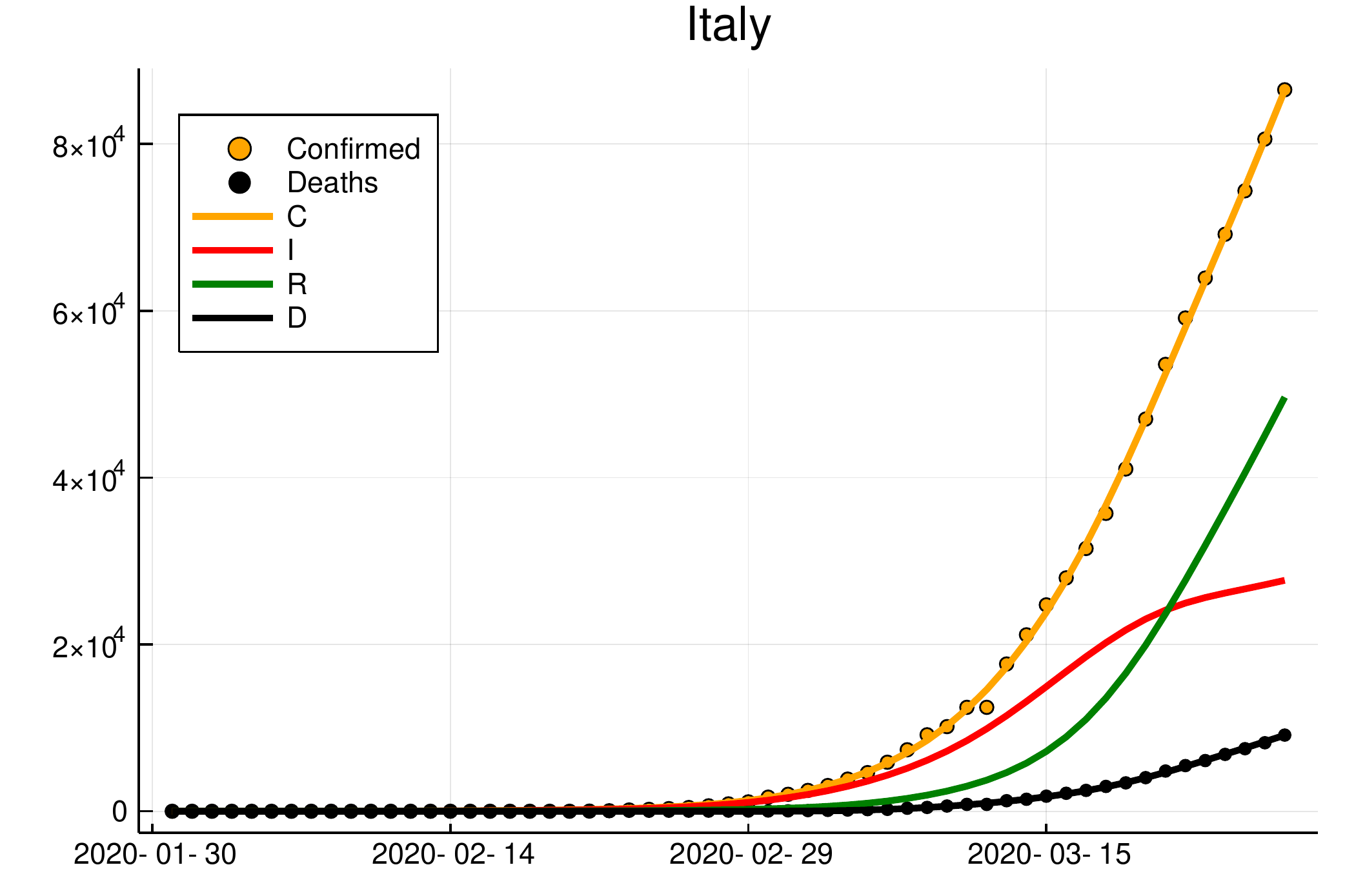}
\caption{Data from Italy (dots) and assimilation (lines)}
\end{figure}

\begin{figure}
\centering
\includegraphics[width=0.75\textwidth]{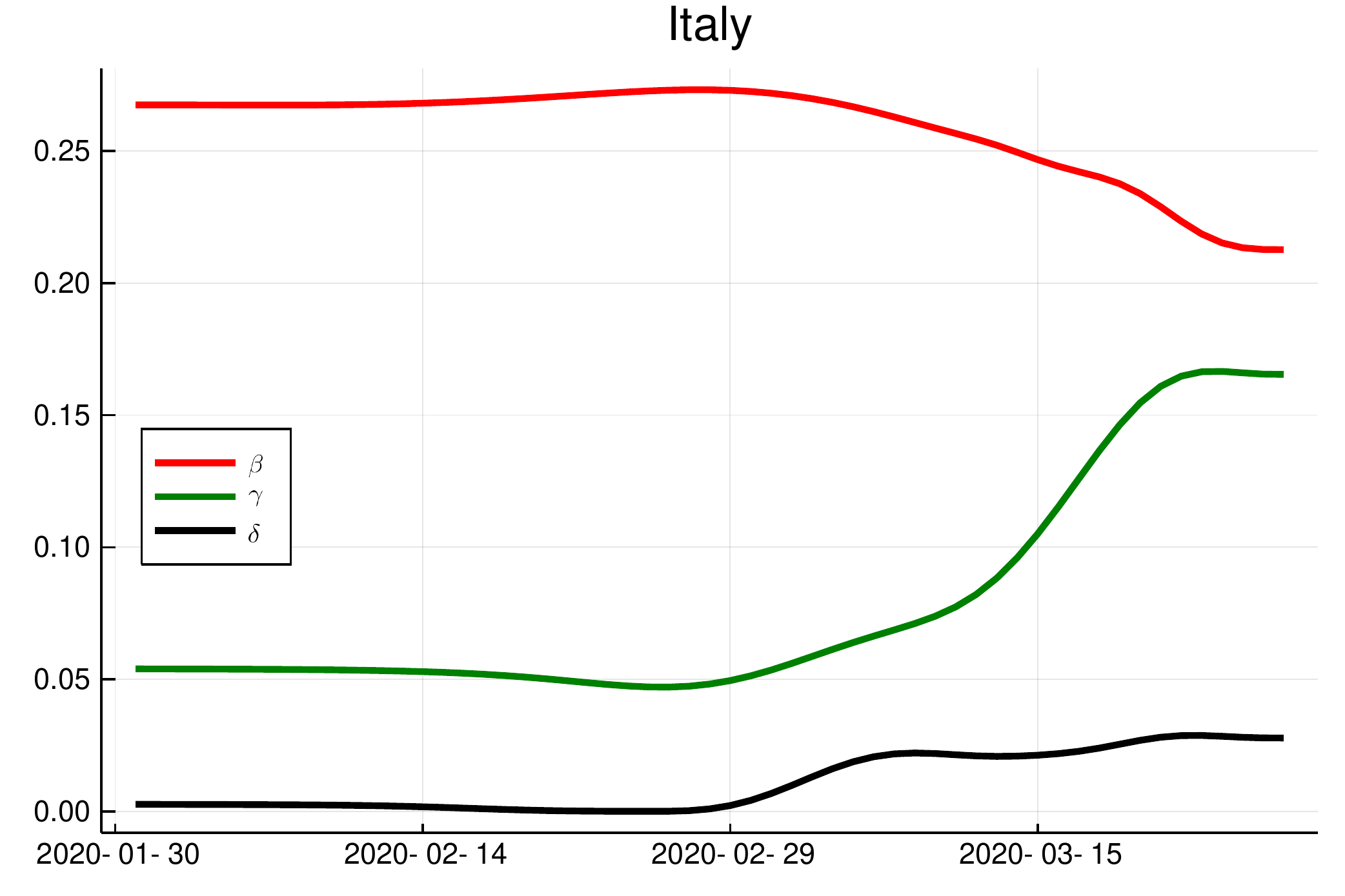}
\caption{Model parameters \(\beta, \gamma, \delta\) for Italy}
\end{figure}
\enlargethispage{24pt}
Italy was the first European country hit by the outbreak. The
assimilated data start at January 30 with the first infected being
reported but the fact that \(\beta\) is already at \(\beta=0.27\)
indicate that the rise, as obvious from the Chinese data and even more
visible in the cases of below, occurred previously. Italy started it's
response in the first days of Mach and the \(\beta\)-value starts to
fall significantly to about \(\beta=0.22\) as of today and seems to have
levelled off there. As already seen in China, the response time of the
\(\beta\) to government action is one month. It is also an important
indicator that the number of infected people is below the ones already
recovered. So far this is only visible in China and Italy, whereas the
cases below are far from there.

\subsubsection{Germany}\label{germany}

\begin{figure}
\centering
\includegraphics[width=0.75\textwidth]{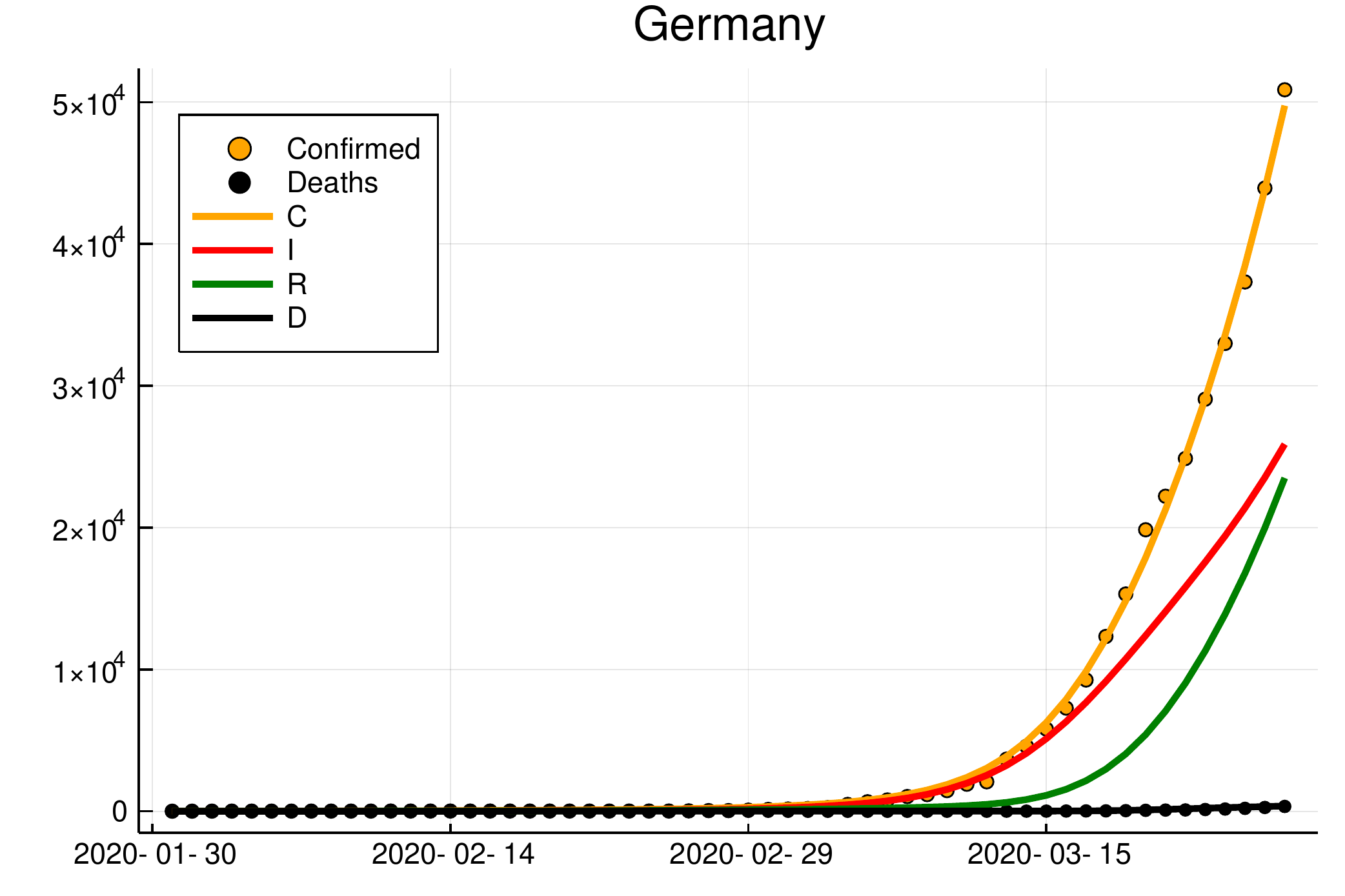}
\caption{Data from Germany (dots) and assimilation (lines)}
\end{figure}

\begin{figure}
\centering
\includegraphics[width=0.75\textwidth]{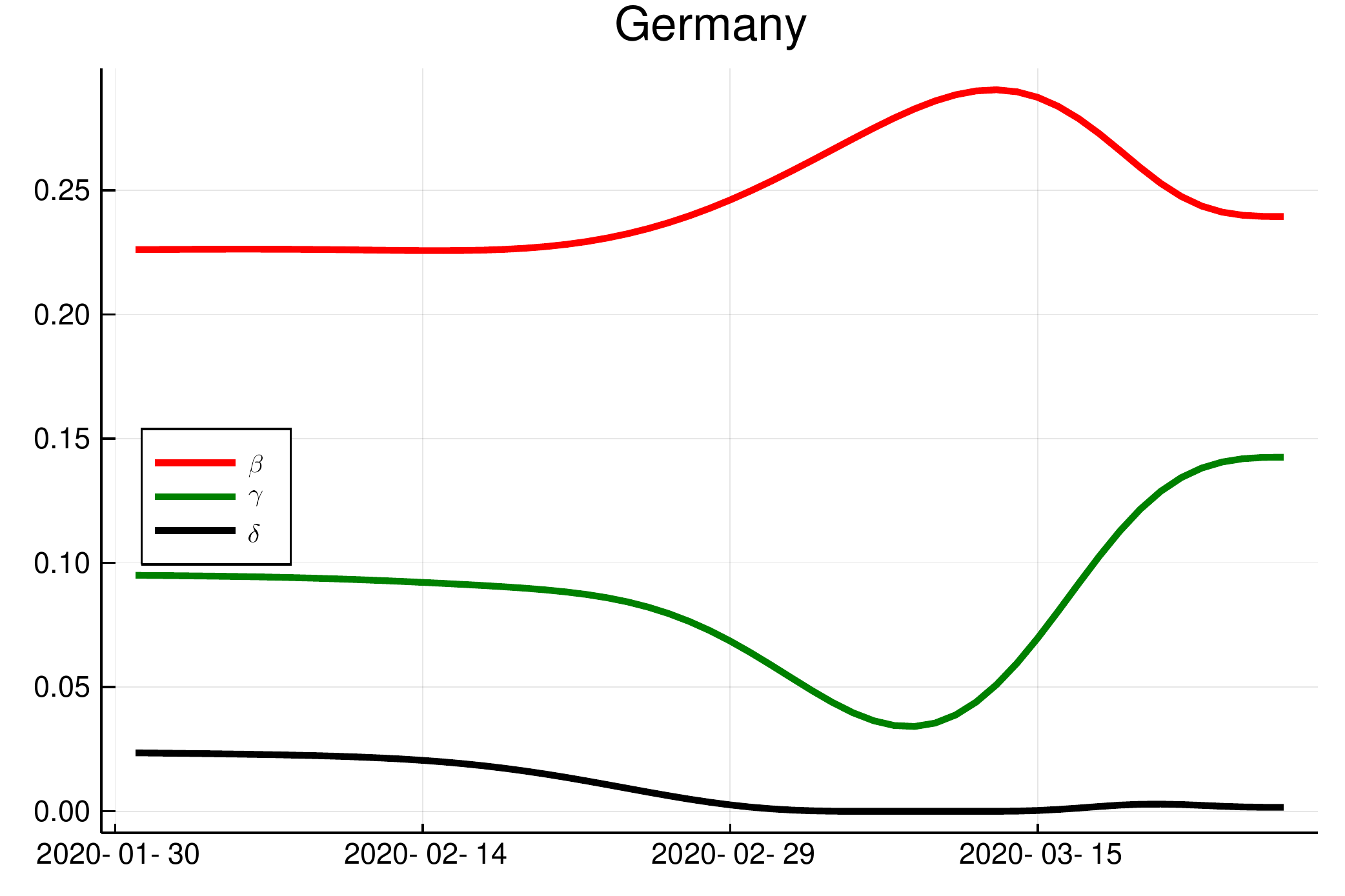}
\caption{Model parameters \(\beta, \gamma, \delta\) for Germany}
\end{figure}

The German data is relatively plain and shows nicely the onset of the
development and the effectiveness of the actions taken. There is an
almost exponential growth and with the naked eye it is hard to tell when
the parameters vary significantly.

A look at the assimilated model parameters show, that there is a
significant increase with a peak around March 15th, starting three weeks
earlier. After March 15th, there is a drop below the originally constant
value within one week. This shows the effectiveness of the government
actions taken mid March. The response time seems to be substantially faster than in
China. The author expects the numbers to decrease further to about
\(\beta=0.2\) within the next two weeks as seen in the case of China and
Italy. As shown in the prognosis for Germany below, the present state
is not sufficient to terminate the outbreak but delays it's peak.
Terminating the countermeasures will likely not \emph{flatten the curve} but
only delay its peak.

\subsubsection{Bavaria}\label{bavaria}

In Bavaria a rise in $\beta$ occurs March 1st.\footnote{Data scraped from \cite{RKI2020}}
Note that at this time
the outbreak is already ongoing with 14 cases reported. Therefore this
rise in $\beta$ significates a substantial acceleration, not the onset.
Note also that the values are subtantially higher than the German-wide
average, where this numbers are included.  The situation is therefore
more severe in Bavaria than on German average. Additional action
surpassing the German-wide restiction was taken 21st of March, which
is clearly visible in a drop of $\beta$ in the assimilation.
\begin{figure}[ht]
\centering
\includegraphics[width=0.75\textwidth]{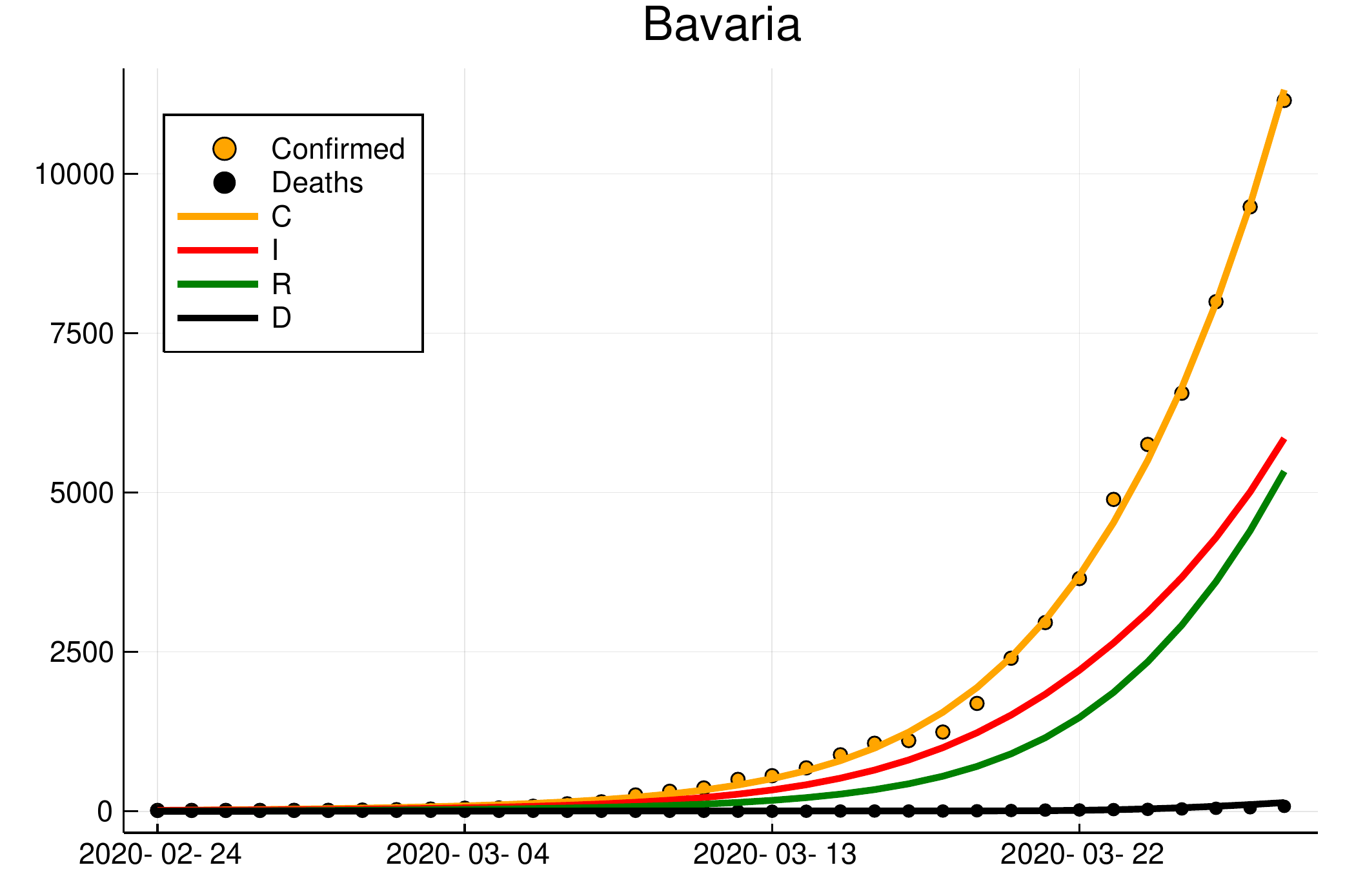}
\caption{Data from Bavaria (dots) and assimilation (lines)}
\end{figure}

\begin{figure}[h]
\centering
\includegraphics[width=0.75\textwidth]{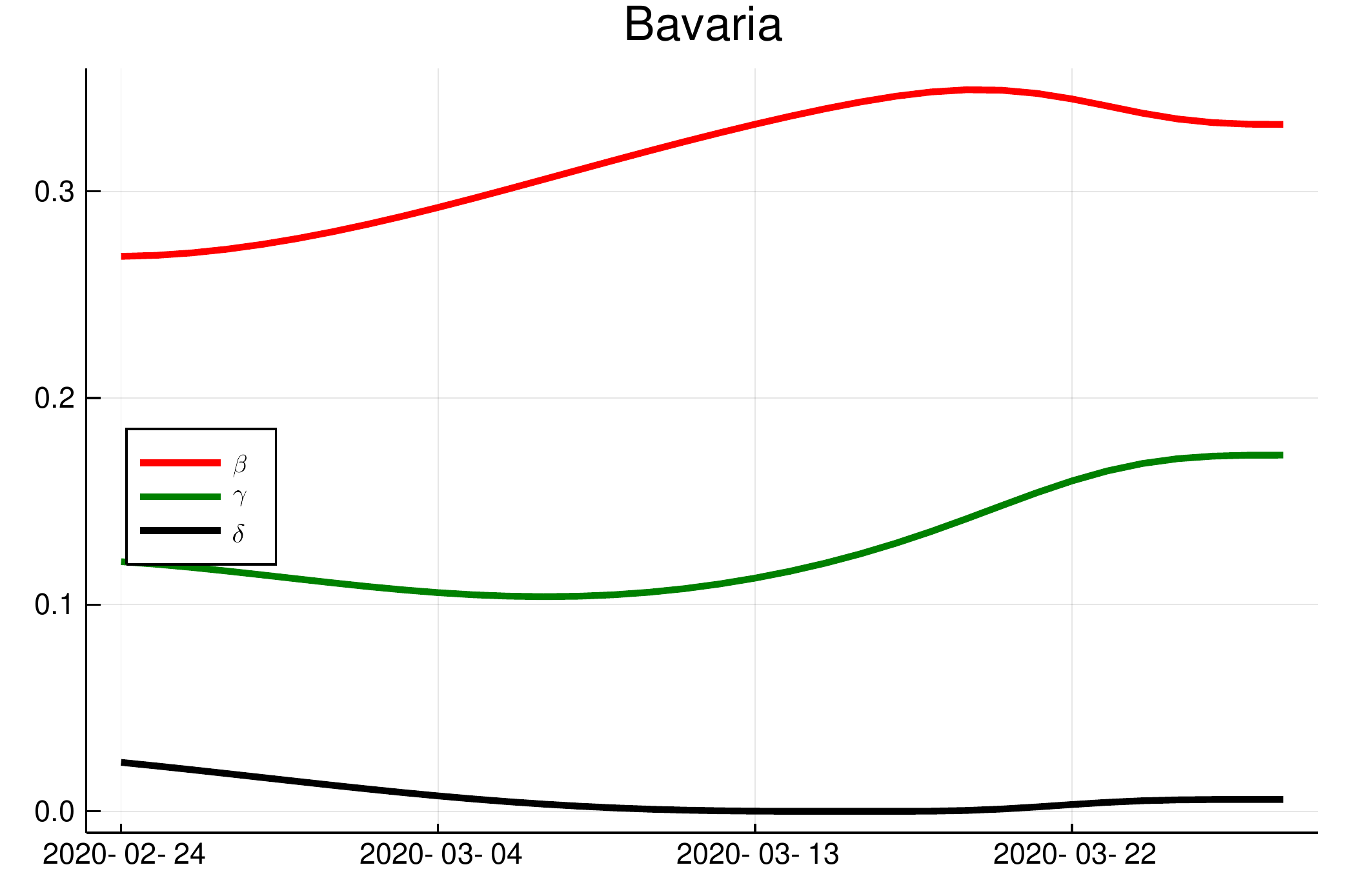}
\caption{Model parameters \(\beta, \gamma, \delta\) for Bavaria}
\end{figure}

\subsubsection{United States}\label{united-states}

The United states have had ongoing cases prior start the start of the
records such that the ramp up of the infection rate to \(\beta=0.24\)
is not visible in the data. However, there was a second surge starting
March 1st which lifted the value up to \(\beta=0.32\) which is well
above what can be seen in European countries but below the Chinese
peak value. There is evidence of a minor sucess in limiting the spread
of the desease. It can also be observed that the recovered count is
still far below the infected. This shows that the US is far from
containing the virus and we have to expect the worst numbers from out
of the US.

\begin{figure}
\centering
\includegraphics[width=0.75\textwidth]{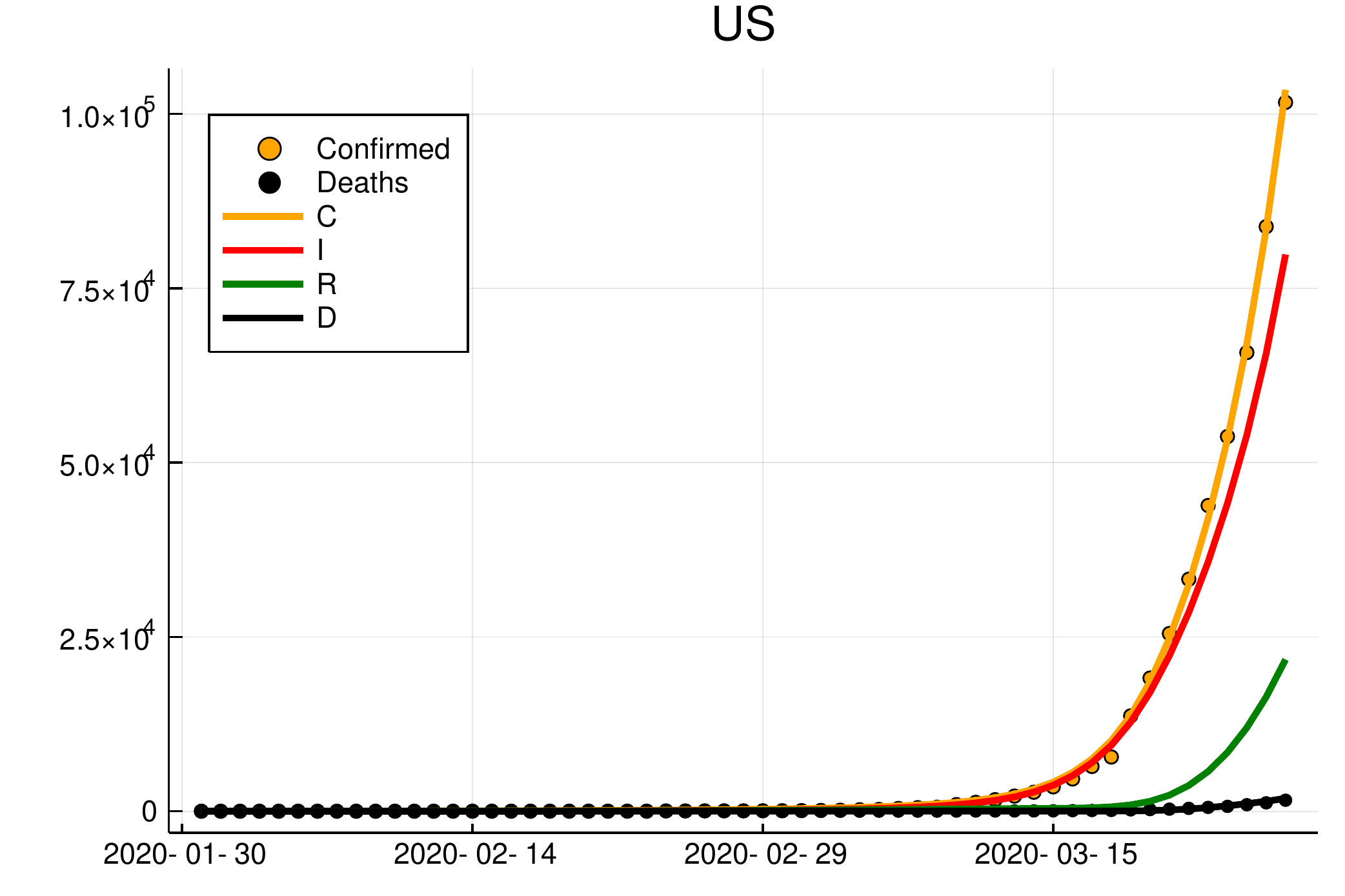}
\caption{Data from the US (dots) and assimilation (lines)}
\end{figure}

\begin{figure}
\centering
\includegraphics[width=0.75\textwidth]{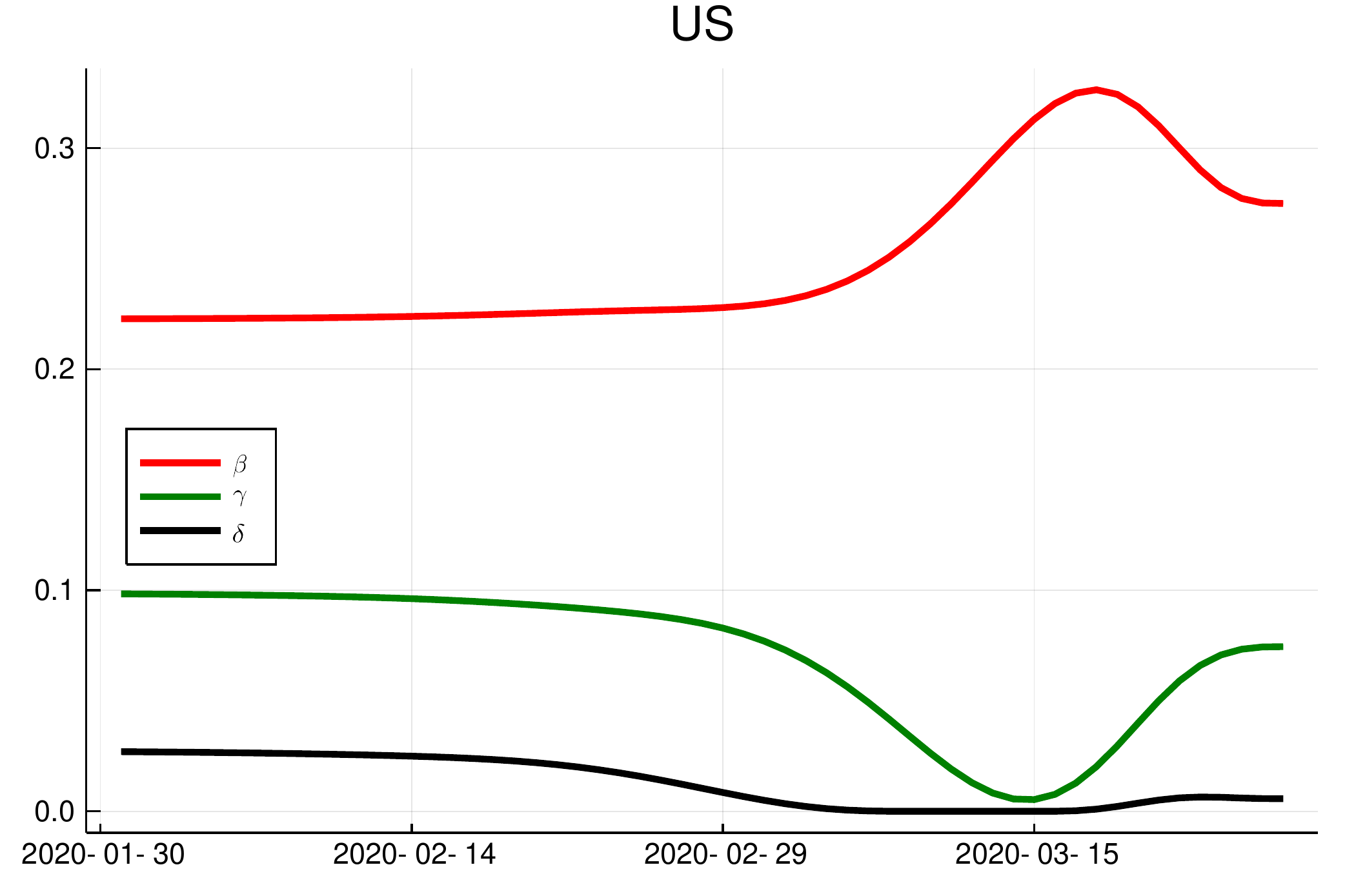}
\caption{Model parameters \(\beta, \gamma, \delta\) for the US}
\end{figure}

\subsubsection{United Kingdom}\label{united-kingdom}

The case of the United Kingdom is akin to the US, but the rise is the
first and hopefully only one, and the growth rate is still
\(\beta = 0.26\). So-far no counter-measurements are visible in the
data. Fortunately the level is substantially lower than the numbers seen
for the US.

\begin{figure}
\centering
\includegraphics[width=0.75\textwidth]{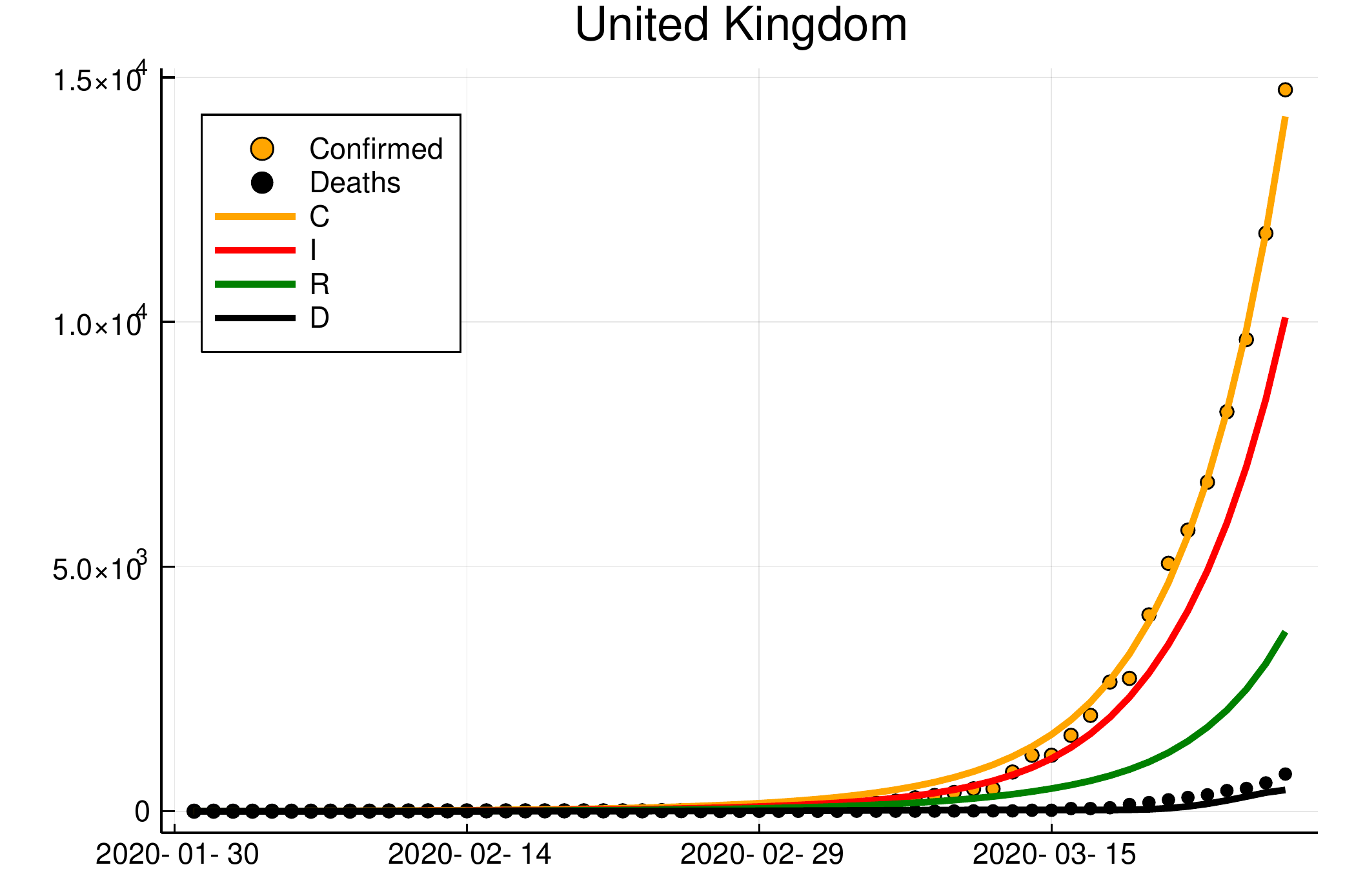}
\caption{Data from the United Kingdom (dots) and assimilation (lines)}
\end{figure}

\begin{figure}
\centering
\includegraphics[width=0.75\textwidth]{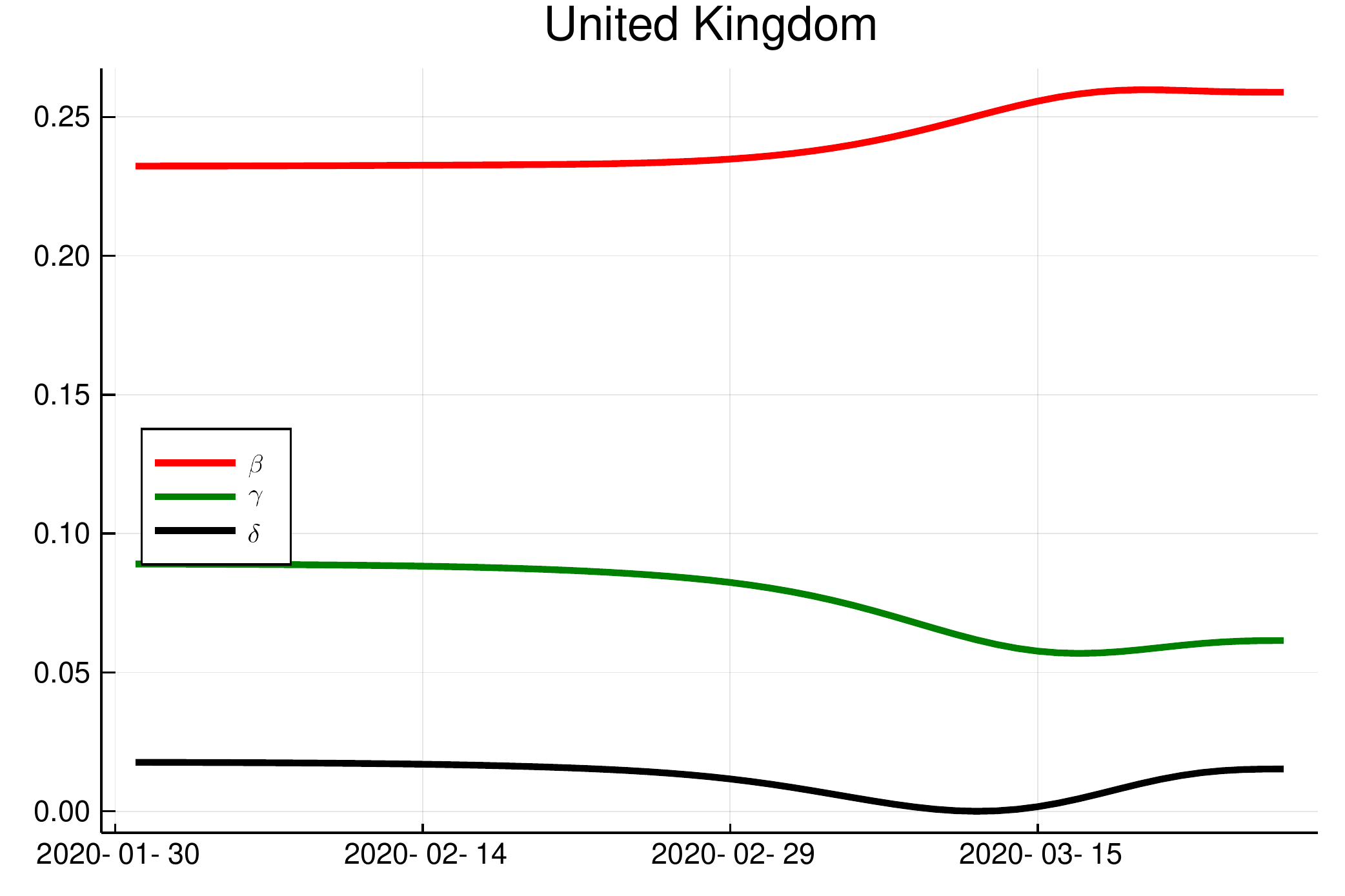}
\caption{Model parameters \(\beta, \gamma, \delta\) for the United
Kingdom}
\end{figure}

\section{Prognosis}\label{prognosis}

Based on the assimilated values of \(\beta\) , \(\gamma\), and
\(\delta\) at the last instance where data is available and the
correspondinng numbers of \(S,I,R,D\), model calculations were made
for the future 400 days. An example is given for the case of Germany.

The result shows, that given the last values of the parameters (which
only partially reflect the effeciveness of the very recent goverment
actions yet) will not limit the outbreak permanently, but delay it until
June. But this is partly good news, because it shows, that the
exponential growth is only slightly positive. It is expected that
numbers within the next days will show the full extent of the goverment
actions.

\begin{figure}[ht]
\centering
\includegraphics[width=1\textwidth]{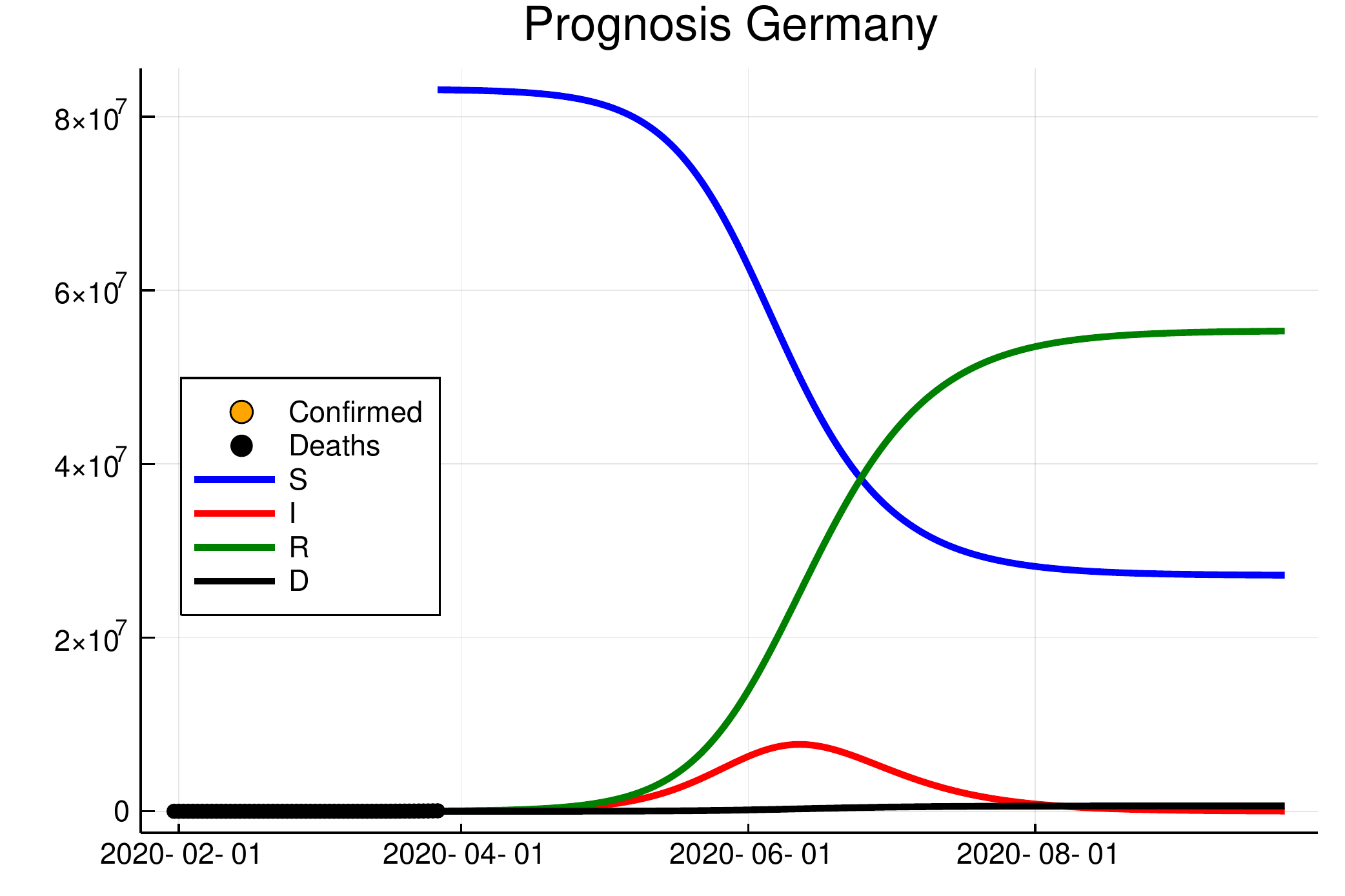}
\caption{Prognosis based on the values for Germany as of March 28th.
Actual cases are to the left and their elacation is not visible in
relation to the total numbers. This indicates the extent of not even a
fully blown outbreak.}
\end{figure}

\section{Conclusions}\label{conclusions}

Data assimilation is able to fit a epidemology model to given data. The
result is the recovery of the full model state \(S,R,I,\) and \(D\), as
well as the model parameters \(\beta, \gamma,\) and \(\delta\). In
particular \(S,R,I, \beta, \gamma,\) and \(\delta\) are difficult to
obtain otherwise.

This gives a unique tool to interpret the data and assess the quality of
govermental action or effect of crucial events like the end of the
holiday season. This was demonstrated in particular by the values of
\(\beta\) and their rise or drop.

Goverment actions, not only in China, but also Italy, Germany, and
Bavaria are effective and can clearly be seen at an early state in
the assimilation results.

The model can be exchanged by a better one as in use by University
Institutions or Goverments. Feedback control is possible to implement,
as well the coupling with for example economic models.  The author is
availiable for requests in that direction.

\section*{Conflict of interest}
 The author declares that he has no conflict of interest.

\bibliographystyle{plain}
\bibliography{literatur}   % name your BibTeX data base

\end{document}